\documentclass[useAMS,usenatbib]{mnras}
\usepackage{amsmath}
\usepackage{graphicx}
\usepackage{txfonts}
\usepackage{natbib}
\usepackage{bm}

\def\apj{ApJ}
\def\apjl{ApJL}
\def\apjs{ApJ Suppl. Ser.}
\def\apss{Astroph.Sp.Sci.}
\def\aap{A\&A}

\def\mnras{MNRAS}
\def\nat{Nature}

\def\physrep{Phys. Rep.}

\def\beq#1{\begin{equation}\label{#1}}
\def\eeq{\end{equation}}
\def\beqa#1{\begin{eqnarray}\label{#1}}
\def\eeqa{\end{eqnarray}}

\def\Eq#1{Eq.~(\ref{#1})} 
\def\eqn#1{~(\ref{#1})}
\def\myfrac#1#2{\left(\frac{#1}{#2}\right)}

\def\comment#1{\relax}

\newcommand{\gcas}{\mbox{$\gamma$\,Cas}}
\newcommand{\lsim}{\raisebox{-.4ex}{$\stackrel{<}{\scriptstyle \sim}$}}
\newcommand{\msim}{\raisebox{-.4ex}{$\stackrel{>}{\scriptstyle \sim}$}}

\title[Enigmatic Be-star $\gamma$~Cas]
{A propelling neutron star in the enigmatic Be-star $\gamma$~Cassiopeia}
\author[Postnov et al.] {
K. Postnov$^{*1}$\thanks{E-mail: pk@sai.msu.ru}, L. Oskinova$^2$, J.M. Torrej\'on$^3$\\
$^1$ Sternberg Astronomical Institute, Moscow M.V. Lomonosov State University,
Universitetskij pr., 13,  Moscow 119234, Russia\\
$^{2}$Institute for Physics and Astronomy, University Potsdam, 
14476 Potsdam, Germany \\
$^{3}$ Instituto Universitario de F\'{\i}sica Aplicada a las Ciencias y las Tecnolog\'{\i}as, Universidad de Alicante, 03690 Alicante, Spain}

\begin{document}

\date{Received ... Accepted ...}
\pagerange{\pageref{firstpage}--\pageref{lastpage}} \pubyear{2015}

\maketitle

\label{firstpage}
\begin{abstract}
The enigmatic X-ray emission from the bright optical star, 
$\gamma$\ Cassiopeia, is a long-standing  problem. 
$\gamma$\ Cas is known to be a binary system consisting of a Be-type star and a low-mass 
($M\sim 1\,M_\odot$) companion of unknown nature orbiting in the Be-disk
plane. 
Here we apply the quasi-spherical accretion theory onto a compact magnetized star 
and show that if the low-mass companion of $\gamma$~Cas is a fast spinning 
neutron star, the key observational signatures of \gcas\ are remarkably well 
reproduced. Direct 
accretion onto this fast rotating neutron star is impeded by the propeller 
mechanism. In this case, around the neutron star magnetosphere a hot shell is formed 
that emits thermal X-rays in qualitative and quantitative agreement  
with observed properties of the X-ray emission from  $\gamma$~Cas. We 
suggest that $\gamma$\,Cas and its analogs constitute a new subclass of 
Be-type X-ray binaries hosting rapidly rotating neutron stars formed in
supernova explosions with small kicks. 
The subsequent evolutionary stage of $\gamma$\,Cas and its 
analogs should be the X\,Per-type binaries comprising low-luminosity 
slowly rotating X-ray pulsars. The model explains the enigmatic X-ray 
emission from \gcas, and also establishes  evolutionary connections between 
various types of rotating magnetized neutron stars in Be-binaries.    
\end{abstract}



\section{Introduction}

The optically brightest Be-star $\gamma$\,Cas (B0.5IVpe) is well seen in the 
night sky by the naked eye even in a big city. It is a well known binary system 
that consists of an optical star with mass $M_{\rm Be}\approx 16M_\odot$ and an 
unseen hot companion with mass $M_{\rm X}\approx 1 M_\odot$. The binary orbital 
plane and the disk of the Be-star are 
coplanar \citep{2000A&A...364L..85H,2002PASP..114.1226M,2007ApJ...654..527G}. Since the discovery 
of X-ray emission from \gcas\ 50\,years ago \citep{1976Natur.260..690M}, its 
enigmatic properties attracted large attention, but so far have
remained unexplained \citep[see recent review by][]{Smith2016}. 

\gcas\ is a prototype of a class of Be-stars that have X-ray luminosities of 
$10^{32}-10^{33}$\,erg\, s$^{-1}$, which is intermediate between those usually 
observed from B-stars with similar spectral types and those of X-ray and 
cataclysmic variable  binaries. The defining feature of X-ray emission from 
the class of \gcas\ analogs is a very hot thermal spectrum with $T\msim 
100$\,MK (or $kT\msim 10$\,keV) and the  presence of fluorescent FeK-line. 
No X-ray pulsations have been detected in \gcas\ analogs. 

After \gcas\ was detected in X-rays, the idea of an accreting neutron 
star (NS) companion had been put forward \citep{White1982} based on
the apparent similarity between the X-ray spectra of 
\gcas\ and the Be X-ray binary (BeXRB) X\,Per that hosts a NS. However, the 
subsequent accumulation and analysis of high-quality multiwavelength  
observations of \gcas\ revealed major difficulties for this model.
For example, \citet{Lopes2006} pointed out that direct accretion onto a NS 
is unlikely because the observed X-ray luminosity of \gcas\ 
(B0.5IVe+NS?, $P_{\rm orb} \approx 204$\,day, $e \lsim 0.03$) would be much lower 
than in X\,Per (O9.5III-IVe+NS, $P_{\rm orb} \approx 250$\,day, $e \approx 
0.1$).
It was also pointed out that the FeK-line is  usually not seen in the X-ray spectra 
of long-period BeXRBs while it is observed in \gcas. Moreover, a non-thermal 
spectral component is typically present in X-ray spectra of BeXRBs, but is
absent in \gcas. \citet{Lopes2006} considered the possibility of 
an unusual accretion regime for a NS, such as accretion onto the NS 
magnetosphere, and concluded that this is unlikely as well. Besides the NS hypothesis, 
at least two other scenarios about the nature of the \gcas\ and its analogs  
have been discussed in the literature -- accretion onto 
a white dwarf and magnetic star-disk interaction.  We will dissuss these
alternative scenarios in Section\,\ref{sec:fin}. 

Despite large discussion in the literature, to our knowledge no quantitative 
model predicting the observed properties of \gcas\ exists so far. In this paper 
we endeavor to develop such model. We derive  quantitative  predictions of 
X-ray emission from a NS in the propeller regime 
embedded in a hot quasi-spherical shell
and compare them with observed properties of \gcas. 

The good agreement between our model and observations lends credence to the 
proposed model 
and allows us to suggest an evolutionary scenario for  
\gcas\ and its analogs. In our new scenario these objects represent natural 
evolutionary stage of binary X-ray systems that experienced mass exchange in the 
past, suffered only a small kick during supernova (SN) explosion of the primary, 
and will evolve to accreting BeXRBs, such as X\,Per, in the future. 

The formation of X\,Per in the context of the standard evolutionary scenario 
for BeXRBs was considered by \citet{Delgado2001}. They found that 
the formation of this  Be+NS system likely involved a quasi-stable and nearly 
conservative transfer of mass from the primary to the secondary. It was 
suggested that the final mass of He star remnant of the primary was less than 
$6\,M_\odot$ and suggested that its supernova explosion might have  been
completely symmetric. Using a Monte Carlo study of 
natal kicks, \citet{Delgado2001} speculated that there may be a substantial 
population of neutron stars formed with little or no kick. 
\citet{Shtykovskiy2005} studied the population of compact X-ray sources in the 
Large Magellanic Cloud, and included the propeller effect to explain the 
observed X-ray luminosity distribution. 

In this paper we expand the standard approach to the high-mass X-ray binary 
evolution \citep[e.g.][]{Pfahl2002} by showing that the quasi-spherical accretion in young 
BeXRBs may be impeded. Our model can have important implications for the physics of 
supernova in massive binary systems and for the formation of double neutron 
star systems in the Galaxy.

The paper is organized as follows: the basic principles of the propeller 
effect at quasi-spherical wind stage are introduced in Section\,\ref{sec:prop}. The properties of a hot 
magnetospheric shell are derived in Section\,\ref{sec:mag}. 
The evolution of the neutron star spin is considered  in Section\,\ref{sec:sd}, 
and its energy balance is estimated in Section\,\ref{sec:en}. A brief summary 
of the model predictions is given in Section\,\ref{sec:pre}, and a comparison of 
the model with observations of \gcas\ is made in Section\,\ref{sec:gc}.
The evolutionary scenario for $\gamma$\,Cas-class of objects is discussed in 
Section\,\ref{sec:ev}. The discussion and conclusion are in 
Section\,\ref{sec:fin}.
 
\section{Propeller effect} 
\label{sec:prop}

Neutron stars are born in core collapses of massive stars. NSs have  
masses of $0.8\,M_\odot\, \lsim\, M_{\rm X}\, \lsim \,
2.5\,M_\odot$ \citep{Lattimer2007}, initially short spin periods $P^*_0\sim 
10-100$~ms, and are strongly magnetized 
with the characteristic dipole magnetic moment $\mu_{30}=\mu/(10^{30}\mathrm{G\,cm}^3)\sim 1$ 
\citep[e.g.][]{pt2012}. 

Some NSs are found in binary systems with a high-mass stellar companion 
of OB or Be spectral type. The OB-stars lose mass via their 
radiatively driven winds, and rapidly rotating Be-stars  
also posses decretion disks \citep{Porter2003,Reig2011}. In the standard 
formation model of high-mass X-ray binaries, a NS gravitationally attracts the 
wind matter outflowing from  its early-type companion 
within the Bondi radius 
\beq{eq:rb}
R_{\rm B}=2GM_{\rm X}/\varv_0^2, 
\eeq
where $\varv_0$ is the NS velocity relative to the wind 
\citep{1972NPhS..239...67V, 1973NInfo..27....3T}.
If the specific angular momentum of the gravitationally captured matter is 
small, the accretion flow is quasi-spherical.
A NS with radius $R_0\approx 
10^6$~cm and accreting matter with density $\rho$ 
at the rate $\dot M_{\rm B}\simeq \piup \rho \varv_0 
R_{\rm B}^2$ could  
gravitationally sustain a luminosity $GM\dot M_{\rm B}/R_0\sim 0.1\dot 
M_{\rm B}c^2$ (where $c$ is the speed of light) \citep{1944MNRAS.104..273B}. 
Hence, if all gravitationally captured matter were able to reach the NS 
surface, a bright X-ray source would appear.

However, for matter to reach the NS surface, it should penetrate through 
its magnetosphere with the characteristic radius $R_{\rm A}\sim 10^8-10^9$~cm, 
defined by the pressure balance between the ambient matter and the magnetic field (see 
Eq.\,\ref{R_m}). 
Besides the magnetospheric barrier, the centrifugal 
barrier can also prevent the matter accretion.  The rigidly rotating 
NS magnetosphere reaches a Keplerian velocity at the distance 
$R_{\rm c}=(GM(P^*)^2/4\piup^2)^{1/3}$, where $P^*$ is the NS spin period. 
Only when the condition $R_{\rm 
c}\le R_{\rm A}$ is met the accretion can start. The corresponding NS 
spin period is then $P_{\rm A}^*\approx 17 
(R_{\rm A}/10^9[\mathrm{cm}])^{3/2}$~s. If $R_{\rm 
A}>R_{\rm c}$ or, equivalently, $P^*<P_\mathrm{A}$), 
the centrifugal barrier at the magnetospheric boundary would 
prevent matter accretion \citep{1975A&A....39..185I, 1986ApJ...308..669S}. 
Such situation is referred to as a `propeller stage'.

The propeller effect has been suggested to operate at low states of  
transient X-ray pulsars 
\citep{2016A&A...593A..16T,2016arXiv160703427L,2016MNRAS.457.1101T}
and has been invoked to explain non-stationary behaviour of supergiant fast 
X-ray transients \citep{2007AstL...33..149G,2008ApJ...683.1031B}. 

\section{The hot magnetospheric shell and its properties}
\label{sec:mag}

In a wind-fed binary system with rapidly rotating NS, 
the propeller effect has an important difference compared to the one operating in the disk-fed systems.
At the propeller stage, the gravitationally captured material 
from the stellar wind of the companion 
will accumulate above $R_{\rm A}$ to form a hot quasi-spherical shell 
extending up to $\sim R_{\rm B}$ \citep{1981MNRAS.196..209D, 
2012MNRAS.420..216S}. The shell can power a gravitational luminosity of
\begin{equation}
L_{\rm X}\approx GM_{\rm X}\dot M_{\rm B}/R_{\rm A} \sim 10^{32}~{\rm 
erg\,s}^{-1}.
\label{eq:Lx}
\end{equation}

To good approximation, the density and temperature distributions in the shell  
can be found from the hydrostatic equilibrium: 
\beq{eq:scalings}
\rho(R)=\rho_{\rm A}(R_{\rm A}/R)^{3/2},\quad T(R)=T_{\rm A}(R_{\rm A}/R)\,
\eeq
where $\rho_{\rm A}$ and $T_{\rm A}$ 
are referred to the near magnetospheric values. The temperature is determined 
by the condition
\beq{eq:tem}
{\cal R}T=\mu_{\rm m}\frac{\gamma-1}{\gamma}GM_{\rm X} \frac{1}{R_{\rm A}},
\eeq 
where ${\cal R}$ is the universal gas constant, $\mu_{\rm m}$ is the molecular 
weight and 
$\gamma$ is adiabatic index of the gas. In the following we will assume 
$\mu_{\rm m}=0.5$ and $\gamma=5/3$. 
The magnetospheric radius is determined from the pressure balance 
at the magnetospheric boundary: 
\beq{R_A}
K_2(B_0^2/8\piup)(R_0/R_{\rm A})^6=\rho_{\rm A}{\cal 
R}T_{\rm A}=\rho_{\rm A}(2/5)(GM_{\rm X}/R_{\rm A})\,,
\eeq 
where the factor $K_2\approx 7.56$ takes into account
compression of a quasi-spherical magnetosphere \citep{1976ApJ...207..914A}. The 
total X-ray luminosity of the shell 
due to bremsstrahlung cooling is
\beq{L_x} 
L_{\rm X}=\int_{R_{\rm A}}^{R_{\rm B}}\epsilon_{\rm br}4\piup r^2dr=
\frac{4\piup K_{\rm br}}{\sqrt{\cal 
R}}\rho_{\rm A}^2\sqrt{\frac{2}{5}\frac{GM_{\rm X}}{R_{\rm A}}}\frac{R_{\rm 
A}^3}{2}
\left(1-\frac{1}{\sqrt{R_{\rm B}/R_{\rm A}}}\right)\,, 
\eeq
where we have used $\epsilon_{\rm br}=K_{\rm br}\rho^2\sqrt{T}$ and the scaling 
laws for the density and temperature in the shell given by \Eq{eq:scalings}. 
The last term in the parentheses can be neglected since usually 
$R_{\rm B}/R_{\rm A}\gg 1$. 
For a given $L_{\rm X}$ and NS  magnetic field $\mu=B_0R_0^3/2$, 
equations (\ref{R_A}) and (\ref{L_x}) can be solved to give
\beq{R_m}
R_{\rm A}\simeq 7.6\times 
10^8[\mathrm{cm}]\mu_{30}^{8/15}L_{32}^{-2/15}(M_{\rm X}/M_\odot)^{-4/15}\,,
\eeq
where $L_{32}=L_{\rm X}/(10^{32}\mathrm{erg\, s}^{-1})$, 
and
\beq{rho_m}
\rho_{\rm A}\approx 4.4\times 
10^{-11}[\mathrm{g\,cm}^{-3}](L_{32}/\mu_{30})^{2/3}(M_{\rm X}/M_\odot)^{1/3}\,,
\eeq
and the electron number density: 
\beq{ne}
n_{\rm e,A}\approx 2.6\times 
10^{13}[\mathrm{cm}^{-3}](L_{32}/\mu_{30})^{2/3}(M_{\rm X}/M_\odot)^{1/3}\,.
\eeq
The temperature at the shell base is 
\begin{eqnarray}
\label{T_A}
&T_{\rm A}=\frac{2}{5}\frac{GM_{\rm X}}{{\cal R}R_{\rm A}}\approx 
27[\mathrm{keV}](R/10^9[\mathrm{cm}])^{-1}\nonumber \\
&\approx 
36[\mathrm{keV}]\mu_{30}^{-8/15}L_{32}^{2/15}(M_{\rm X}/M_\odot)^{19/15}\,.
\end{eqnarray}
This high temperature justifies the use of bremsstrahlung radiative losses from 
the shell. 

The volume emission measure of the shell is given by
\begin{eqnarray}
\label{ME}
&{\rm EM}=\int_{R_{\rm A}}^{R_{\rm B}}n_{\rm e}^2(r)4\piup r^2dr=4\piup 
n_{\rm e,A}^2R_{\rm A}^3\ln\myfrac{R_{\rm B}}{R_{\rm A}}\nonumber \\
&\approx 3.7\times 10^{54}[\mathrm{cm}^{-3}]
\mu_{30}^{4/15}L_{32}^{14/15}(M_{\rm X}/M_\odot)^{-2/15}\ln\myfrac{R_{\rm 
B}}{R_{\rm A}}\,.
\end{eqnarray}
In the context of a NS coplanar with the Be-star disk, given the slow equatorial disk 
wind velocities $\varv_0\sim 10^7$~cm s$^{-1}$, we find $R_{\rm B}/R_{\rm A}\sim 100$ 
(see Eq.\,\ref{eq:rb}), and the EM can be $\sim 
10^{55}[\mathrm{cm}^{-3}]$ and even higher. 

\section{Evolution of the neutron star spin}
\label{sec:sd}

The propeller effect can be important in astrophysical context provided that its duration is sufficiently
long compared to the life time of the Be-star (several million years).
The propeller effect operates only for fast 
rotating NSs. Therefore, to evaluate for how long accretion may be inhibited and 
a hot magnetospheric shell can be supported by the wind from the optical star, 
one should consider the spin evolution of a non-accreting NS.

The transfer of angular momentum in magnetospheric shells around quasi-spherically accreting NSs 
was considered in more detail by \citet{2012MNRAS.420..216S}. It was found,
in particular, that in such shells a nearly iso-momentum angular velocity distribution is 
established, $\omega(R)=\omega_{\rm m}(R_{\rm A}/R)^2$,
where $\omega_{\rm m}$ is angular velocity of matter at the magnetosphere. However, when 
accretion onto the NS is centrifugally prohibited, the angular momentum transport by viscous forces 
through the surrounding convective shell  leads to the angular 
velocity distribution $\omega(R)=\omega_{\rm m}(R_{\rm A}/R)^{7/4}$  
\citep[see Appendix A6][]{2012MNRAS.420..216S}). 

From Eq.\,(51) and (52) presented in \citet{2012MNRAS.420..216S}, at the 
stage with no accretion, the braking torque applied to NS from the surrounding shell is   
\beq{sd}
I\dot\omega_*=-\frac{49}{4}\omega_B^2\myfrac{R_{\rm B}}{R_{\rm A}}^{7/2}\piup C 
\rho_{\rm A} R_A^5,
\eeq 
where $I$ is the NS  moment of inertia, $\omega_B=2\piup/P_{\rm orb}$ is 
the binary orbital frequency, $C\gtrsim 1$ is a 
numerical coefficient that determines 
turbulent viscosity through the Prandtl law and hence viscous stresses in the 
convective shell. Plugging into \eqn{sd} the density distribution in 
the shell, $\rho_{\rm A}=\rho_{\rm B}(R_{\rm B}/R_{\rm A})^{3/2}$, as given by
\eqn{rho_m} and expressing it through the X-ray luminosity $L_{\rm X}$, 
\Eq{sd} for the braking torque can be rearranged to
\beq{sd1}
I\dot\omega_*=-49\omega_{\rm B}^2R_{\rm B}^3 C\myfrac{R_{\rm A}L_{\rm 
X}}{GM_{\rm X}\varv_0}\,.
\eeq 
The characteristic NS spin-down time in this regime thus becomes:
\begin{eqnarray}
\label{tsd}
&t_{\rm sd}\equiv\frac{I\omega_*} {I\dot\omega_*}=
\frac{I\omega_*}{49\omega_{\rm B}^2R_{\rm B}^3 C}
\myfrac{GM_{\rm X}\varv_0}{R_{\rm A}L_{\rm X}}\nonumber \\
&\approx 2\times 
10^5[\mathrm{yr}]\myfrac{P_*}{1\mathrm{s}}^{-1}\myfrac{P_{\rm 
orb}}{100\mathrm{d}}^2\myfrac{\varv_0}{100\mathrm{km\,s}^{-1}}^7
\myfrac{R_{\rm A}}{10^9\mathrm{cm}}^{-1}L_{32}^{-1},
\end{eqnarray}
(here the constant $C$ was set to unity.) 

As can be seen from Eq.\eqn{tsd}, the NS spin-down time is extremely 
sensitive to the NS  velocity relative to the disk wind (as $\sim \varv_0^7$) 
and to binary orbital period (as $\sim P_{\rm orb}^2$). It can be made much longer 
than the characteristic spin-down time at the propeller stage during the 
disk accretion onto NS  with the same magnetospheric radius, 
$t_{\rm sd,d}\approx (I\omega_*)(\mu^2/R_{\rm A}^3)^{-1}\sim 
10^5[\mathrm{yr}] 
(P_*/1\mathrm{s})^{-1}\mu_{30}^{-2}R_{\rm A,9}^3$.

Given the significant time duration estimated by Eq.\eqn{tsd}, it is obvious that 
propelling NSs surrounded by hot quasi-spherical shells can 
be present among low-luminosity non-pulsating high-mass X-ray binaries. 
Their X-ray spectral properties as summarized below in Section \ref{sec:pre} and their long orbital 
binary periods 
can be used to distinguish them from, for example, faint hard X-ray emission from 
magnetic cataclysmic variables \citep[e.g.][]{2016ApJ...825..132H}.

\section{Heating of the magnetospheric shell}
\label{sec:en}

In the case of a  `supersonic propeller' \citep{1981MNRAS.196..209D}, additional 
source of the shell heating is provided by the mechanical energy flux from 
the spinning-down NS \citep[sometimes referred to as 'magnetospheric 
accretion', see][]{1986ApJ...308..669S}. 
Multiplication of Eq.\eqn{sd} by the angular velocity difference at the 
magnetospheric boundary, $(\omega_*-\omega_{\rm m})$,
gives the influx of the mechanical energy into the shell at the propeller 
stage. 
Clearly, once $\omega_{\rm m}\to \omega^*$ 
during the NS  spin-down, the mechanical energy supply to the shell should vanish. 
However, even in this case the shell can be kept hot due to the 
gravitational energy release given by Eq.\eqn{eq:Lx}.  

To provide X-ray luminosity at the level 
$L_{\rm X}\sim 10^{32}-10^{33}$\,erg\,s$^{-1}$ and assuming 
$R_{\rm A}\sim 10^9$\,cm, the gravitational capture rate of stellar wind matter 
must be $\dot M_{\rm B}\sim 10^{15}-10^{16}$\,g\,s$^{-1}$ or 
$10^{-10}-10^{-11}\,M_\odot$\,yr$^{-1}$. Using Eqs.\eqn{eq:scalings} and 
\eqn{rho_m}, 
it can be shown that  the 
wind density at the outer boundary of the shell near $R_{\rm B}$ is sufficient 
to provide the required mass accretion rate. 

The above estimates are done using simple spherically symmetric considerations, 
which may be violated in complex regions near $R_{\rm B}$. However, it is 
important to note that the observed X-ray luminosity is mostly 
determined by the Bondi-Hoyle rate, $\dot M_{\rm B}\sim \rho_{\rm B} 
R_{\rm B}^2 \varv_0$. 
Expressing it through the density near the shell base eliminates 
the ill-known value of the wind velocity:
\begin{eqnarray}
\label{dotMB}
&\dot M_{\rm B}\sim (1/4)\piup\rho_{\rm B} R_{\rm B}^2 v_0 
\simeq (1/4)\piup\rho_{\rm A} R_{\rm A}^{3/2}R_{\rm B}^{1/2}\varv_0\nonumber \\
&\approx 6\times 
10^{15}[\mathrm{g\,s}^{-1}]\mu_{30}^{2/15}L_{32}^{4/5}(M_{\rm 
NS}/M_\odot)^{-2/5}
\end{eqnarray}
(here the factor 1/4 takes into account density jump in the strong shock near 
$R_{\rm B}$). Thus, our basic considerations should not be strongly affected by 
the complicated flow details at $R_B$ and provide robust first-order estimates.

The radiation cooling time of the hot plasma near the 
base of the shell is rather short, 
$t_{\rm cool}\sim 2\times 10^{11}[s]T^{1/2}n_{\rm e}^{-1}\sim 1000$~s. 
The temperature gradient in the quasi-spherical shell turns out to be 
superadiabatic \citep{2012MNRAS.420..216S}, indicating the presence of 
convection. To avoid rapid cooling, the convection should lift 
up a hot parcel of gas faster than it radiatively cools down, i.e. the 
condition $t_{\rm cool}>t_{\rm conv}$, where $t_{\rm conv}=R_{\rm A}/v_{\rm 
conv}$ is the characteristic time of convective overturn near the shell base, 
should be met. 

The convective velocity is $v_{\rm conv}=\epsilon_{\rm c} c_s$, where 
$c_s=\gamma{\cal R}T$ is the adiabatic sound velocity, $\epsilon_c\le 1$. 
Plugging $R_{\rm A}$ and $T_{\rm A}$ from Eq.\eqn{R_m} and Eq.\eqn{T_A}, we
obtain for the condition $t_{\rm cool}>t_{\rm conv}$
\beq{tcool>tconv}
(\mu_{30}L_{32})^{2/5}<115\epsilon_c(M_{\rm X}/M_\odot)^{28/15}\,, 
\eeq   
which is easily satisfied even for small convective velocities. 
Convection  initiates turbulence,  and 
the hot thermal plasma in the quasi-spherical shells around NS magnetospheres 
should show signs of turbulent velocities with
$\varv_{\rm turb}\sim \varv_{\rm conv}\sim 1000$\,km\,s$^{-1}$.

\section{Brief summary of the model predictions}
\label{sec:pre}

Lets us summarize the basic properties of the hot magnetospheric shell 
supported by a propelling NS in circular orbit in a binary system around a Be-star. 
The model predicts the following observables: 

\smallskip\noindent
i) the system emits optically thin multi-temperature thermal radiation with the 
characteristic temperatures above $\sim 10$\,keV, high plasma densities $\sim 10^{13}$~cm$^{-3}$ 
and emission measures $\sim 10^{55}$~cm$^{-3}$;

\smallskip\noindent
ii) the typical X-ray luminosity of the system  is $\sim 10^{33}$\,erg\,s$^{-1}$;

\smallskip\noindent
iii) no X-ray pulsations are present and no significant X-ray outbursts are 
expected in the case of a coplanar circular orbit with the Be-disk;

\smallskip\noindent
iv) the hot shell is convective and turbulent, therefore the observed X-ray emission lines 
from the optically thin plasma should  be broadened up to $\sim 1000$\,km\,s$^{-1}$;

\smallskip\noindent
v) the typical size of the hot shell is $\sim R_B \lsim R_\odot$;

\smallskip\noindent
vi) the cold material, such as the Be-disk in the vicinity of 
the hot shell, should give rise to fluorescent FeK-line; 

\smallskip\noindent
vii) the life-time of a NS in the propeller regime in binaries with long orbital periods 
can be  $\sim 10^6$\,yrs, hence such systems should be observable among faint X-ray binaries in the Galaxy. 

\section{Observed X-ray properties of $\gamma$\,Cas can be explained by the 
presence of a propelling neutron star}
\label{sec:gc} 
 
The model predictions outlined in Sect.\,\ref{sec:pre} match very well 
the properties of \gcas\ deduced from observations. We suggest that 
the low-mass companion of $\gamma$ Cas 
can be a NS in the propeller stage. 
The NS orbits the Be-star in almost circular orbit coplanar with the Be-disk. 
The Be-disk is contained within the Roche lobe of the Be-star ($\sim 310 R_\odot$)
\footnote{The Be-disk size in \gcas\  as measured by the infrared interferometry \citep{2007ApJ...654..527G} and inferred from 
emission lines spectroscopy \citep{1988A&A...189..147H,1992A&AS...95..437D} is about two times as small, apparently because these observations 
sample mostly the densest innermost parts of the disk-like wind outflow; millimeter photometry 
indeed suggests a larger disk radius, $\sim 33 R_*$, which is close to the Roche lobe size \citep{1991A&A...244..120W}.}, and
the NS gravitationally captures matter from the slow Be-disk equatorial wind, which is not limited by the Roche lobe of the Be-star.
The scale-height of the disk outflow in \gcas\  is $H_{\rm disk} =0.04 R_*$ 
\citep{2011MNRAS.416.2827M}, comparable to the aspect ratio of the 
Bondi radius, which is sufficient to realize the quasi-spherical accretion. 
The lack of periodic pulsations, as well as properties of the hot 
thermal plasma measured from the analysis of  X-ray observations ($kT_{\rm 
hot}\sim 20$\,keV and  $n_{\rm e}\sim 10^{13}$\,cm$^{-3}$, $\varv_{\rm 
turb}\sim 1000$\,km\,s$^{-1}$)
\citep[e.g.][]{2010A&A...512A..22L,2012ApJ...750...75T,Shrader2015}, which 
challenge all previously proposed scenarios of X-ray emission from $\gamma$~Cas 
\citep{Smith2016}, are naturally expected in a hot magnetospheric shell around a 
propelling NS (Eqs.\,\ref{eq:Lx}, \ref{rho_m}, \ref{T_A} and 
Section \ref{sec:pre}). 

The propeller model explains both the gross physical parameters of hot plasma 
in $\gamma$\,Cas and matches the properties of its X-ray 
variability. Indeed, a hot convective shell above the NS magnetosphere should display 
the time variability in a wide range that depends on the characteristic sound speed, $c_s$, which 
is of the order of the free-fall time, $t_{\rm ff}$. The shortest time-scale is
$t_{\rm min}\sim R_{\rm A}/c_{\rm s}\sim R_{\rm m}/t_{\rm ff}(R_{\rm m})\sim 
R_{\rm m}^{3/2}/\sqrt{2GM}\sim$ a few seconds, while the longest time scale is 
$t_{\rm max}\sim 
R_{\rm B}/t_{\rm ff}(R_{\rm B})\sim 10^6[\mathrm{s}] (\varv_0/100 
\rm{km\,s}^{-1})^{-3}\sim$ a few 
days. These are indeed the typical time scales of the X-ray variability observed 
in $\gamma$\,Cas \citep{2010A&A...512A..22L}. 

Typically, Be-type stars display significant time variability in the optical that 
is produced by changes in the mass-loss rates due to stellar pulsations 
and possible viscose instabilities in the circumstellar disk 
\citep{Baade2016}. In complex systems consisting of a pulsating Be-star, decretion
Be-disk and a NS, 
one can expect the characteristic time delay between any changes in the 
stellar mass-loss rate and Be-disk (usually observed in the optical) and the 
response of the hot magnetospheric shell (usually observed in the X-rays) 
to these changes. 

The dynamics of disks in binary systems is complicated, especially in the 
case of large mass ratio, as in the Be+NS case  \citep[see, e.g., recent 
2D-simulations ][]{Orazio2016}. 
In addition, in a Be+NS  system, the decretion Be-disk is subjected to a number 
of perturbations and resonances which truncate the outer disk edge within 
the Roche lobe of the Be-star \citep{2001A&A...377..161O}. The Roche lobe 
radius of the NS is
\beq{eq:rlns}
R_{\rm L}(M_{\rm X})/a\simeq 0.49 \myfrac{M_{\rm 
X}}{M_{\rm Be}+M_{\rm X}}^{1/3}\approx 0.2
\eeq
where $a$ is the binary semi-major axis,
and we used $M_{\rm X}/M_{\rm Be}\approx 1/16$ for $\gamma$ Cas.     
The characteristic time delay  
is then determined by the free-fall time inside the NS's Roche lobe, 
\beq{tffRL}
t_{\rm ff}(M_{\rm X})\approx \sqrt{\frac{R_{\rm L}^3(M_{\rm X})}{2GM_{\rm 
X}}}=\frac{P_{\rm 
orb}}{2\piup}\sqrt{\frac{R_L(M_{\rm X})}{2}}\sqrt{\frac{M_{\rm 
Be}+M_{\rm X}}{M_{\rm X}} }\,,
\eeq
where we have used 3d Kepler's law $\omega_B^2=G(M_{\rm Be}+M_{\rm X})/a^3$. 
Plugging values for $\gamma$ Cas immediately yields $t_{ff}\sim 40$~days, which 
is close to the time lag between optical  and X-ray 
variability observed in $\gamma$ Cas \citep{2015ApJ...806..177M}. 

Changes in the absorption column density are also expected due to 
perturbations in the cold disk and wind induced by the NS on short and long 
time-scales \citep{2011MNRAS.416.2827M,2012A&A...540A..53S,Hamaguchi2016}.
Note also that the single power-law spectrum $P(f)\sim 1/f$  over the wide 
frequency range from 0.1\,Hz to $10^{-4}$\,Hz, as derived for the 
X-ray time variability in $\gamma$\,Cas \citep{2010A&A...512A..22L}, is common 
for accreting X-ray binaries  and is thought to 
arise in turbulent flows beyond the magnetospheric boundary 
\citep{2009A&A...507.1211R}.

Future works on sophisticated multi-dimensional numeric hydrodynamic models of 
Be-stars, 
their disks and companion NSs  will provide more insight into the 
complex interactions in such systems. Even in the non-relativistic cases of 
Be-binaries, the hydrodynamic models show that the shape of the disk is 
affected by the secondary \citep{Panoglou2016}. Such modeling is 
required to explain the full range of variability observed in \gcas\ and its 
analogs.

\section{$\gamma$\,Cas analogs in the context of evolution of massive binaries}
\label{sec:ev}

During the last decade, about a dozen Be-stars sharing similar characteristics 
with \gcas\ were discovered \citep[][and references therein]{Smith2016}. 
It became clear that \gcas\ is not a peculiar object but a representative 
of a whole class of objects. We propose the following evolutionary scenario for 
\gcas\ analogs.

All $\gamma$\,Cas-type stars were likely formed through  similar evolutionary 
channels. Consider, for example, the standard evolutionary scenario of a
massive binary system with almost equal initial masses $M_{\rm p}\sim 10-11\,M_\odot$ 
and $M_{\rm s}\sim 8\,M_\odot$ separated by 20 solar radii. After the main-sequence stage,  
the primary overfills its Roche lobe  
and transfers a significant amount of matter and angular momentum 
to the secondary. The hydrogen envelope of the primary is  
stripped off during the mass transfer to leave the 
naked helium-rich primary remnant with the mass $M_{\rm 
He}\sim 0.1 M_\mathrm{p}^{1.4}\simeq 3\,M_\odot$. The secondary mass increases up to $M_{\rm Be} \sim 
15\,M_\odot$, and the star acquires rapid rotation. Rotating at 
nearly break-up velocity, the secondary is now observed as an early Be-type star 
with the surrounding disk. Then the helium star explodes as an electron-capture SN 
(ECSN) and produce a NS with mass $M_\mathrm{X}\sim 1 M_\odot$ \citep[e.g.][]{Postnov2014,Moriya2016}. Supernovae of this type  
naturally produce NSs with low-velocity kicks \citep{Podsiadlowski2004, 
2010NewAR..54..140V}.

After the ECSN with low kick, the newly born NS remains in the plane of 
the Be decretion disk and stays in an orbit with low-eccentricity $e=(M_\mathrm{He}-M_\mathrm{X})/(M_\mathrm{Be}+M_\mathrm{X})\sim 0.1$ 
but moves to a somewhat wider separation (in \gcas\ the putative NS is at 
$\sim 35\,R_\ast$ from the Be-star). Importantly, due to low kick velocity, the NS orbits the Be-star 
in the Be-disk plane. Be-stars have high dense equatorial winds, the orbital velocity of NS relative to the 
Be-disk wind can be low, which potentially provides an efficient quasi-spherical accretion. 
However, the high spin of the young NS inhibits accretion of matter captured from the low-velocity wind of the Be-star. 
Instead, the NS is embedded in a hot shell, which we presently observe in X-rays.

The hot shell mediates 
the angular momentum transfer from the NS magnetosphere and can prolong the propeller 
stage of the NS up to several $10^5$~years or even longer. 
Therefore, in binaries with long orbital periods, such as $\gamma$\,Cas, 
the duration of the propeller stage can be comparable to the life-time of the 
Be star (see Eq.\,\ref{tsd}). Thus, a sizable fraction of Be X-ray binaries 
in the propeller stage, i.e. $\gamma$\,Cas analogs, should exist in the Galaxy, as indeed 
observed.

With time, the NS slows down. Once its spin period becomes such that 
corotation radius equals the magnetospheric radius the accretion may begin. 
The NS spin period 
reaches equilibrium (hundreds of seconds in the quasi-spherical wind, \cite{2012MNRAS.420..216S}) 
at which torques acting on the NS, on 
average, vanish. With the beginning of accretion, the NS will become a
slowly rotating X-ray pulsar. It should remain in a circular orbit in 
the plane of the Be decretion disk and have a relatively large orbital 
separation. The observational manifestation of such post-$\gamma$\,Cas 
system is X\,Per -- a slowly rotating X-ray pulsar at the stage of quasi-spherical settling 
accretion with moderate X-ray luminosity \citep{2012MNRAS.423.1978L}. 

\section{Discussion and conclusion} 
\label{sec:fin}

\gcas\ and its analogs constitute a well observed and well studied class of 
objects. The vast literature on the subject was recently reviewed by 
\citet{Smith2016}, and we refer the interested reader to that comprehensive 
review. Besides detailed multi-wavelengths observations, the three often invoked 
scenarios for the nature of \gcas\ are also discussed in depth by 
\citet{Smith2016}, and we only briefly repeat these scenarios here. 

Chronologically, the first hypothesis explaining the enigmatic X-ray 
properties of \gcas\ was that of an 
accreting NS \citep{White1982}. Yet, the improvement in X-ray  
spectroscopy and timing led to the realization that the X-ray properties of 
\gcas\ are very different from accreting NSs in other Be X-ray binaries.   

A different commonly invoked scenario is the accretion 
onto a white dwarf \citep{1995A&A...296..685H,2002A&A...382..554A}. 
\citet{Smith2016} disfavor the white dwarf hypothesis because of the too 
high observed X-ray luminosity of \gcas. Such X-ray luminosity is
difficult to achieve without increasing the supply of matter from 
the Be-disk by discrete ejections, and these are not frequent enough. 
\citet{Lopes2006} considered accretion onto a magnetic white dwarf. 
Our estimates show that a propelling magnetic white dwarf as the  
companion in $\gamma$\,Cas is disfavored since, with the typical white dwarf 
magnetic moment of a few $\times 10^{32}-10^{33}$\,G\,cm$^3$, the 
magnetospheric radius given the observed low X-ray luminosity would be too 
large (see Eq.\,\ref{R_m}) and gas temperature too low (see Eq.\,\ref{T_A}) 
to match the values derived from X-ray spectroscopy. Recently 
\citet{Hamaguchi2016} invoked the white dwarf hypothesis to explain the observed 
changes in the X-ray spectral hardness ratio by the presence of absorbers that 
occult the hot spots on the white dwarf surface.  Their estimates show that the 
X-ray luminosity 
and the absorber densities  can be explained using plausible assumptions about 
densities and velocities in the stellar wind and disk. However, the lack of coherent 
pulsations from the accreting white dwarf remains unexplained. 

The third scenario explaining \gcas\ and its analogs does not relate to binarity. 
Instead it suggests a magnetic star-disk interaction. In this picture, 
the entanglement and stretching of magnetic loops from the stellar surface with 
the disk magnetic field lines lead to reconnection events that 
accelerate particles. These electron streams bombard the stellar surface. 
The thermalisation leads to the X-ray emission. Albeit some estimates on 
the thermalisation efficiencies are made, this scenario is entirely 
phenomenological at present and is lacking any predictive power 
\citep{Smith2016}. Note that recently Schoeller et al. (submitted to A\&A) 
attempted to detect magnetic fields in some \gcas-analogs 
using spectropolarimetric observations, but no evidence for magnetic fields have been found.   

In this paper we propose a novel explanation of the enigmatic class 
of \gcas\ analogs. Using the well established physical models 
of propelling NSs 
surrounded by hot convective magnetospheric 
shells in wind-fed accreting systems, we derive quantitative predictions  and compare them 
with the key observational properties of \gcas. This comparison leads us to 
conclude that a propelling 
NS in \gcas\ matches theoretical predictions very well. Moreover, the 
existence of young fast spinning NS companions (propellers) to early Be-stars 
in circular and coplanar orbits is a natural consequence of the standard evolutionary scenario 
of massive binary stars and ECSN models. Hence, the synergy between the stellar evolution and accretion 
theories predicts the existence of \gcas\ and its analogs.     
 
\section{Acknowledgements} 
The authors acknowledge ISSI (Bern) for 
hospitality. We thank the anonymous referee for useful comments and suggestions that helped to improve this paper. The work of KP is supported by RSF grant 16-12-10519. JMT 
acknowledges the research grant ESP2014-53672-C3-3-P and LO 
acknowledges the DLR grant 50 OR 1508. 

\bibliographystyle{mnras}
\bibliography{gammacas}

\end{document}